\documentclass[useAMS,usenatbib]{mn2e}
\usepackage{courier}
\usepackage{multirow}
\usepackage{times}
\usepackage{upgreek}
\usepackage{graphicx}
\usepackage{epstopdf}
\usepackage{amssymb,amsmath}
\usepackage[section]{placeins}

\title[A detection threshold in {\it Kepler} data]{A detection threshold in the amplitude spectra calculated from {\it Kepler} data obtained during K2 mission}
\author[A.S. Baran]
{A.S. Baran $^1$\thanks{E-mail:sfbaran@cyf-kr.edu.pl},
C.Koen $^2$
and
B. Pokrzywka $^1$
\\
$^{1}$ Uniwersytet Pedagogiczny, Obserwatorium na Suhorze, ul. Podchor\c{a}\.zych 2, 30-084 Krak\'ow, Polska\\
$^{2}$ Department of Statistics, University of the Western Cape, Private Bag X17, Bellville, 7535 Cape, South Africa
}
\begin{document}

\date{}

\pagerange{\pageref{firstpage}--\pageref{lastpage}} \pubyear{2010}

\maketitle

\label{firstpage}

\begin{abstract}
We present our analysis of simulated data in order to derive a detection threshold which can be used in the pre-whitening process of amplitude spectra. In case of ground-based data of pulsating stars, this threshold is conventionally taken to be four times the mean noise level in an amplitude spectrum. This threshold is questionable when space-based data are analyzed. Our effort is aimed at revising this threshold in the case of continuous 90-day {\it Kepler} K2 phase observations. Our result clearly shows that a 95\% confidence level, common for ground observations, can be reached at 5.4 times the mean noise level and is coverage dependent. In addition, this threshold varies between 4.8 and 5.7, if the number of cadences is changed. This conclusion should secure further pre-whitening and helps to avoid over-interpretation of spectra of pulsating stars observed with the {\it Kepler} spacecraft during K2 phase. We compare our results with the standard approach widely used in the literature.
\end{abstract}


\section{Introduction}
\label{introduction}
Data of pulsating stars contain noise (of a variety of origins) and an intrinsic signal. Very often the signal is periodic.
which creates a coherent signal easily picked up by a Fourier transform. In an amplitude spectrum (a scaled square root of the traditional periodogram), each signal is represented by a peak which is located at a frequency corresponding to the pulsation period, and reaches a height close to signal amplitude. By contrast, noise is uncorrelated, hence it will not be in phase over the course of observations, leading to a random distribution of amplitudes over frequencies.  Peaks associated with a real signal are selected, based on amplitude spectra, and they are pre-whitened from time-series data. Such pre-whitening is continued until all peaks with amplitudes satisfying {\it certain condition} have been removed.

This {\it certain condition} (hereafter: a detection threshold) indicates a significance level of a peak. The detection threshold is commonly adopted to be a {\it signal-to-noise} (S/N) ratio $\geq$\,4; S denotes the height of the peak in question while N is the average noise level in an amplitude spectrum. Such a condition was claimed to be a reasonable limit for ground-based data by {\it e.g.} \cite{breger93} or the Hubble Space Telescope data by \cite{kuschnig97}, and many authors have subsequently used this limit.

The detection of a periodic signal hidden in noise has always been a challenge in astronomy. While there exists numerous papers dealing with this issue, we specifically bring a few of them to a reader's attention. \cite{scargle82} reported on the efficiency of detection by means of the {\it periodogram} in the case of unevenly spaced times series data. He provided, through the {\it false alarm probability}, a simple estimate of the significance of the height of a peak in the power spectrum. It became a widely accepted tool for astronomers to distinguish between a signal and noise. \cite{horne86} considered a different normalization of the periodogram and showed that only use of the total variance leads to exponential behavior of the probability distribution function, and validated the resulting estimates of the false alarm probability. \cite{kuschnig97} analyzed Hubble Space telescope data and derived a criterion, given a specific probability, to predict an upper limit for peaks in the amplitude spectrum of time series data. Detection sensitivity of the oscillation modes has been provided by \cite{kjeldsen92}, who analyzed CCD ground based observations. \cite{reegen07} provided a tool for reliable computation of significance levels in the frequency domain, based on the false-alarm probability associated with a peak in the amplitude spectrum. A discussion of methods used for determining the significance of peaks in periodograms of time series has also been undertaken by \cite{frescura08}. It is worth noting that all this work have been done under the null hypothesis {\it Are data consistent with  pure noise?}

When detecting peaks in amplitude spectra two types of errors exist. The first is  detection of a spurious peak and can be related to the cumulative distribution function of noise. The second one is the non-detection of the true periodic signal, which is associated with the cumulative distribution function of the noise and the signal. It should be stressed that both distribution functions depend on the noise distribution as well as on data sampling and windowing. Since it is extremely difficult to derive an analytical formula linking the detection threshold with its confidence level, data simulation are used. Work which is directly relevant to the contents of this paper was included in Master's degree project of \cite{miedzinska99} (hereafter EM). The goal of the project was to find new pulsating subdwarf B stars and the simulations were done to eliminate spurious signal which may exist in data.

\section{Methodology}
\label{method}
EM generated Gaussian noise with a given standard deviation. Then a sinusoidal signal with fixed period of 700\,c/d was added. The range of amplitudes of the signal ranged between S/N\,=\,2.5 to 6.0 in different numerical experiments. A detection threshold for the simulated data was established by counting datasets in which a peak at 700\,c/d was the highest. The number of detections provide the confidence level of finding a specific peak to be real in a dataset. EM showed that a peak with an amplitude of S/N\,=\,4 corresponds to 95\% confidence level. This level was adopted to be high enough to consider a peak to be real, hence, a detection threshold of S/N\,=\,4 was confirmed.

The simulations described above were performed on data characteristic of ground-based observations. Such data usually have short cadences while their coverage is either short or patchy. Fairly often a number of different sites, using different photometric systems, are used to achieve a longer coverage. The {\it Kepler} spacecraft has opened a new way to collect time-series data of pulsating stars. The coverage is almost continuous while data are of unprecedented quality and taken by means of one optical setup. The only non-uniformity comes from different silicons (or positions within the central silicon) used to collect data of a specific object.

In the case of {\it Kepler} data the large number of cadences in time-series data increases the probability of identifying a spurious frequency in an amplitude spectrum of the entire dataset. This argument was frequently given by authors presenting analyses of {\it Kepler} data \citep[][among others]{baran12} and to be on the safe side, they considered peaks with S/N close to 4 to be tentative. To dispense with these doubts we undertook an analysis of evenly spaced simulated data, based on the methodology presented by EM, to estimate a detection threshold for representative datasets obtained with the {\it Kepler} spacecraft, limited to the coverage achievable during K2 phase.

\section{Data analysis}
\label{data}
We used \textsc{Python} to simulate our datasets. We generated Gaussian noise with only one sinusoidal signal, of the form $A \cos(2 \pi t f+\varphi)$, injected. We expect that the S/N required for a satisfactory confidence level may occur at similar values, as compared to ground-based data. Therefore, the range of values of the amplitude A was used, to cover a range of S/N ratios from 1 to 7, at
intervals of S/N\,=\,0.5. In each simulated dataset, the frequency {\it f} and phase $\varphi$ of the signal were random values in [0,734.07]\,c/d and [0,2$\pi$]\,rad, respectively.

We analyzed data sets described by three different noise standard deviations, characteristic of three pulsating subdwarf B stars observed during the {\it Kepler} mission but limited to the arbitrarily chosen one quarter of coverage, which is comparable with the expected K2 coverage. The standard deviations were: 13.5\,ppt (S1), 3\,ppt (S2) and 0.5\,ppt (S3), where ppt denotes {\it parts per thousand}. The values were adopted from data for KIC\,2991403 (K=17.14$^{\rm mag}$), KIC\,2697388 (K=15.39$^{\rm mag}$) and KIC\,9472174 (K=12.26$^{\rm mag}$), respectively for S1, S2 and S3. We considered short cadence data sampled every 58.85\,sec.

For each standard deviation and each injected amplitude we simulated 1000 time-series datasets. The data contained either 135\,000 (S1,S3) or 130\,000 (S2) points.

The form of the spectrum used was
\begin{eqnarray}
F(f)&=&\sqrt{\frac{2}{N}} \left \{
\left [ \sum_t (y_t-\overline{y}) \cos 2\pi f (t-\tau) \right ]^2 C^{-1}+ \right .\nonumber\\
&&\left .
\left [ \sum_t (y_t-\overline{y}) \sin 2\pi f (t-\tau) \right ]^2 S^{-1}
\right \}^{1/2}\nonumber
\end{eqnarray}
where
$$C=\sum_t \cos^2 2\pi f(t-\tau) \;\;\;\;\;\;\;\;
S=\sum_t \sin^2 2\pi f(t-\tau)$$
and 
$$\tau = \frac{1}{4\pi f}\tan^{-1} \left [ 
\left (\sum_t \sin 4\pi f t \right ) \left / 
\left (\sum_t \cos 4\pi f t \right ) \right . \right ]$$
The spectrum $F(f)$ conveniently estimates the amplitude, rather than power, associated with frequency {\it f}.

We calculated an amplitude spectrum from 0 to the Nyquist frequency (734.07\,c/d for short cadence data) with a step of 0.00121\,c/d. If the amplitude spectrum is only calculated in the Fourier frequencies, then the different amplitude values are uncorrelated, and the distribution of the spectrum maximum is easily calculated. In practice, in order to make sure important peaks are not missed, the spectrum is oversampled. This means that the amplitude spectrum values in different frequencies are correlated, and the distribution of extreme values changes; this is well known in the theory of extreme value distributions, e.g. \citep{kotz00,castillo05}. In practice, the ad-hoc S/N$\geq$\,4 has therefore been used. This paper assesses the reliability of this criterion. Of course, in practical applications the true S/N is unknown, and the usual practice of using the estimated value in place of the true value would be followed. Since the datasets under consideration are very large, the estimated S/N should be quite accurate.
%(It is also intuitively easy to see that this will have an influence: if strong peaks are oversampled, then the number of very large values will be increased).

To decrease the computation time we used a Fast Fourier Transform algorithm. Then, in each amplitude spectrum we searched for a peak within 0.01\,c/d from the frequency of the signal injected and we checked if that peak was the highest in the entire amplitude spectrum. Finally, we counted the amplitude spectra meeting that condition. In Figure\,\ref{histogram1} we present a relation between the S/N ratio of the injected signal and the {\it fraction of datasets with correct frequency determinations}.

We stress that our work differs from the usual approach in which simulations are done under the null hypothesis H0, {\it there is no signal in the data}. Since there is sampling variation in both the noise spectrum, and the observed signal spectrum (due to interaction between noise and signal spectra), spurious large noise-induced peaks, or too low signal-related peaks may be produced. Both spectra can contribute to the largest peak not corresponding to the signal frequency, therefore rejection of H0 does not guarantee that the {\it correct} frequency has been identified. We are working under the alternative hypothesis H1, {\it there is a signal at the peak frequency}, which allows us to derive the probability that the signal is correctly identified in the data.

\section{Results}
\label{results}
\begin{figure}
\includegraphics[width=\hsize,height=200mm]{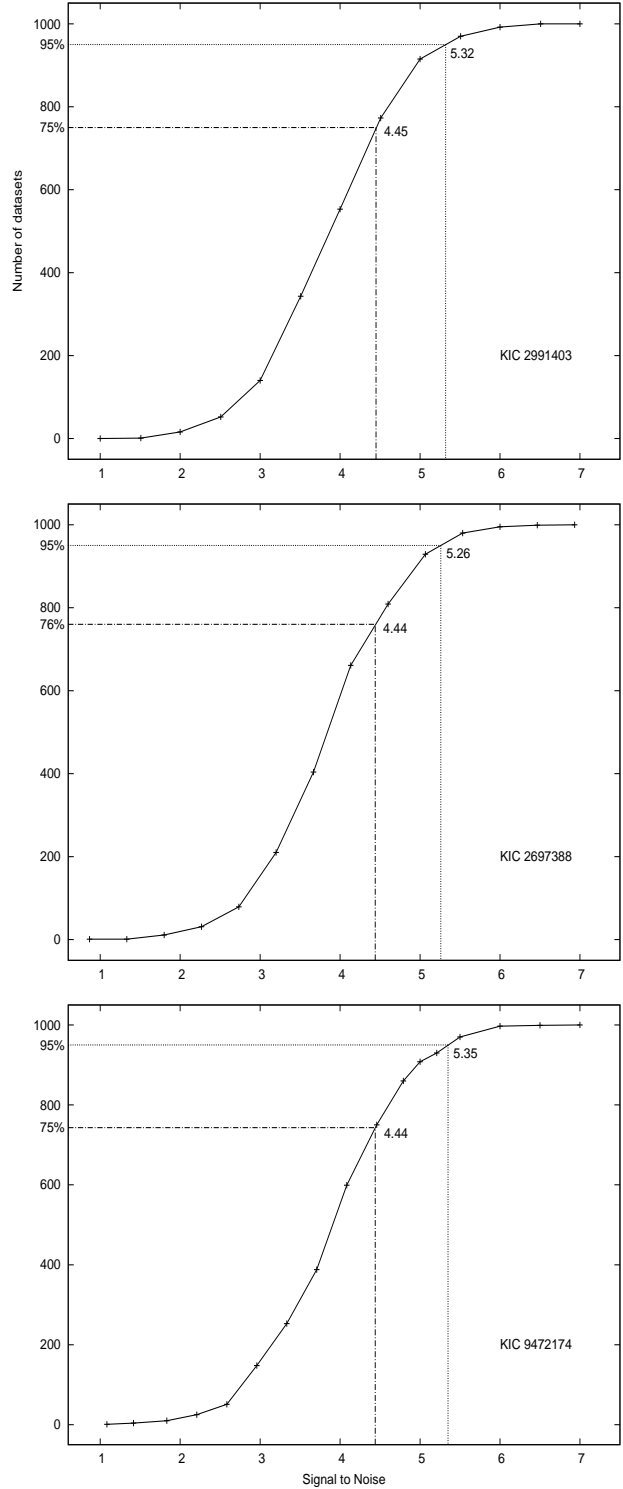}
\caption{Power functions for three different values of the noise standard deviation. 95\% confidence levels and the S/N are marked with dotted lines. For comparison, the dot-dashed and double-dotted lines mark 95\% points in case of null hypothesis H0 of pure noise. See Section\,\ref{results} for details.}
\label{histogram1}
\end{figure}

As could be expected, the {\it number of datasets} with correct frequency detections increases with increasing amplitude of the signal. The curves in Fig\,\ref{histogram1} have the shapes of logistic functions, i.e. typical of cumulative distribution functions. If we adopt a probability of 95\% (950 counts) as high enough to consider a detection to be reliable, then we can accept a S/N of at least 5.4 to be a reasonable detection threshold. The 95\% confidence level varies with $\sigma$ only marginally. This is expected since $\sigma$ only scales all the amplitudes in the amplitude spectrum. 

\begin{figure}
\includegraphics[width=\hsize,height=200mm]{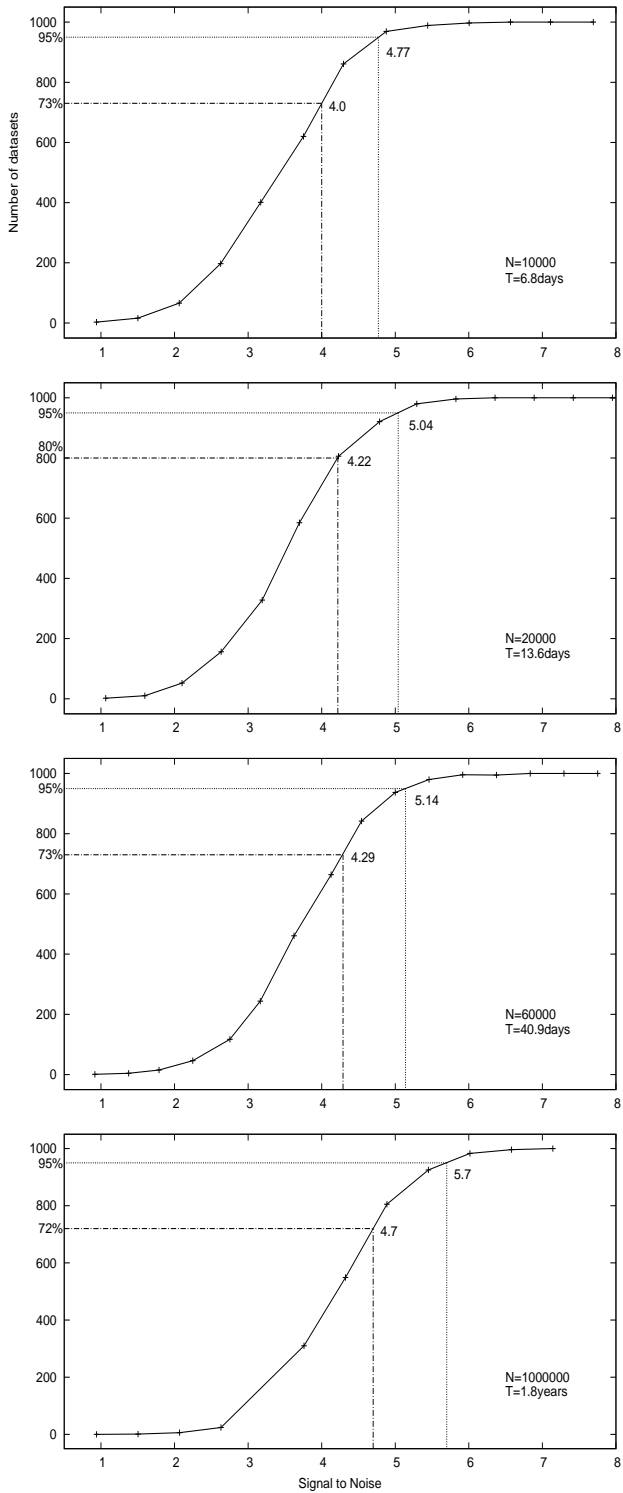}
\caption{Same as Figure\,\ref{histogram1} but for four different data coverages.}
\label{histogram2}
\end{figure}

We repeated our simulations varying the number of data points for the arbitrarily chosen fixed $\sigma$\,=\,3. We confirm that the detection threshold changes with the number of points in a sample. In the case of N=10\,000 cadences the 95\% level is achieved at S/N=4.77; for N=20\,000 cadences we find S/N=5.04; for N=60\,000 we obtained S/N=5.14; for N=1\,000\,000 we derived S/N\,=\,5.7. These results show that changing the data coverage of {\it Kepler} data modifies the detection threshold (Figure\,\ref{histogram2}). This is in agreement with our expectation that increasing the number of points in a time-series increases the probability of a spurious detection.

In conclusion, if we assume that a 95\% confidence level is high enough to distinguish between spurious and true signals, then we consider a S/N\,=\,5.4 (aka 5.4$\sigma$ limit) to be a reliable and safe condition. This detection threshold is appropriate for time-series data obtained with the {\it Kepler} spacecraft during K2 phase. 

For comparison purposes, we calculated the detection threshold under the null hypothesis H0 based on simulations of pure noise. We calculated 1000 time-series datasets of pure noise for the same values of $\sigma$ and N as in Section\,\ref{data}. Then, we ordered the simulation results in terms of the ratio (maximum peak value/mean noise level), extracted from each individual simulation, and determined
the 95\% points over all simulations. These percentiles are indicated by dot-dashed lines in Figs.\,\ref{histogram1} and \ref{histogram2}. For N\,=\,135\,000, the 95-th percentile obtained from simulations of pure noise equals 4.45; it corresponds to a 75\% confidence level in simulations of noise+signal. The thresholds and confidence levels for other N values can be read from Fig.\,\ref{histogram2}.

%N=10000 - mean noise=53.3ppm
%N=20000 - mean noise=37.6ppm
%N=60000 - mean noise=21.8ppm
%N=1000000 - mean noise=5.3ppm

\section*{Acknowledgments}

The work was supported by Polish National Science Centre under project No.\,UMO-2011/03/D/ST9/01914. CK acknowledges funding by the South African National Research Foundation. The authors thank the anonymous referee for suggestions which helped improve the paper.

\bibliography{abaran}

\label{lastpage}

\end{document}